\documentstyle[12pt]{article}
\title{\bf Dynamics of the Universal Confining String Theory on 
the Loop Space}
\author{D.V.ANTONOV \thanks {On leave of absence from the 
Institute of Theoretical and Experimental Physics (ITEP); 
supported by Graduiertenkolleg 
{\it Elementarteilchenphysik}, DFG-RFFI, Grant 436 RUS 113/309/0 
and by the INTAS, Grant No.94-2851. 
E-mail: antonov@pha2.physik.hu-berlin.de
} 
\\
{\it Institut f\"ur Elementarteilchenphysik, Humboldt-Universit\"at 
zu Berlin,}\\
{\it Invalidenstrasse 110, D-10115, Berlin, Germany}}
\date{}
\begin{document}
\maketitle
\vspace{1cm}
\centerline{\bf {Abstract}}
\vspace{3mm}
Starting with the representation of the Wilson average in the Euclidean 
4D compact QED as a partition function of the Universal Confining 
String Theory, we derive for it the corresponding loop equation, 
alternative to the familiar one. 

In the functional momentum representation the obtained equation  
decouples into two independent ones, which describe the dynamics 
of the transverse and longitudinal components of the area derivative 
of the Wilson loop. At some critical value  
of the momentum discontinuity, which can be determined from a certain 
equation, the transverse component does not propagate.

Next, we derive the equation for the momentum Wilson loop, where on the 
left-hand side stands the sum of the squares of the momentum 
discontinuities, multiplied by the loop, which describes its  
free propagation, while the right-hand side describes 
the interaction of the loop with the functional vorticity tensor current. 

Finally, using the method of inversion of the 
functional Laplacian, we obtain for the Wilson loop in the 
coordinate representation a simple Volterra type-II linear integral 
equation, which can be treated perturbatively.
\newpage
{\large \bf 1. Introduction}

\vspace{3mm}
Recently a new progress in the string representation of the confining 
phase of gauge theories was achieved by the development of the 
so-called Universal Confining String Theory (UCST)$^{1}$, whose partition 
function is nothing else than the Wilson loop in the 3D compact QED. It 
was argued in Ref.1 that the summation over the branches of the 
multivalued action of the UCST corresponds to the summation over the 
string world sheets, which is a new step in the understanding of the 
connection between gauge fields and strings.

In Ref.2 the action of the UCST in 4D was derived by performing 
the exact duality transformation, and the low-energy limit of this theory 
with the $\theta$-term included was investigated. This low-energy 
expansion was already discussed in Ref.1 (in the low-energy limit the 
Wilson loop may be evaluated exactly), and in Ref.2 this program was 
elaborated out, so that the string tension of the Nambu-Goto term and 
the coupling constant of the rigidity term were calculated explicitly. 
The sign of the obtained rigid string coupling constant is negative, 
which means that this low-energy string effective action is stable 
(see for example Ref.3). Also the string $\theta$-term appeared in 
the low-energy expansion due to the field-theoretical one.

The most important property of the UCST lies in the observation, proposed 
in Ref.1, that this string theory describes all the gauge theories in such 
a way that the contribution to the partition function of the UCST, 
going from the group $U(N)$, enters with the 't Hooft factor $N^{-\chi}$, 
where $\chi$ is the Euler character of the string world sheet. This 
conjecture seems to bee justified by proving the fact that the partition 
function of the UCST satisfies loop equations$^{4,5}$ modulo contact 
terms, which can't be analized since their structure depends upon the 
topology of the string world sheet. 

In this letter we shall use the representation of the Wilson average 
in the Euclidean 4D compact QED as a partition function of the UCST 
in order to derive for it the corresponding loop equation, 
alternative to the familiar one. Thus, the 
equation, which will be obtained, is nothing else than the equation of 
motion of the partition function of the UCST, written in the loop space. 
The low-energy limit of the UCST, i.e. the case when $\frac{H_
{\mu\nu\lambda}^2}{m^6e^6\exp\left(\frac{const}{e^2}\right)}\ll 1$, 
where $H_{\mu\nu\lambda}$ is the strength tensor of the Kalb-Ramond field, 
$m$ is the mass of this field, and $e$ is the 
coupling constant, corresponds to an obvious simplification of this loop 
equation.

Next, we shall perform the functional Fourier transformation$^{6,7}$ and 
rewrite this equation in the momentum representation, where the functional 
differential operator, acting onto the area derivative of the 
Wilson loop, may be easily inverted, and the resulting equation  
decouples into a system of two independent ones, which describe the 
dynamics of the transverse and longitudinal components of the  
area derivative of the loop. We shall see that at some critical value  
of the discontinuity of the momentum loop, which can be determined 
from the corresponding equation, the transverse part of the loop area 
derivative does not propagate. In the low-energy 
limit of the UCST this equation possesses only a spurious solution, and 
therefore the transverse component always propagates in this limit. 
After that, we shall 
bring the momentum loop equation to such a form, that on 
its L.H.S. will stand the sum of the squares of the momentum 
discontinuities, 
multiplied by the momentum loop, which is known to describe the 
free propagation of the 
loop$^{6}$, while the R.H.S. of this equation will describe 
the interaction of the loop with the functional vorticity tensor current. 

Finally, we shall return to the coordinate representation and make 
use of the method of inversion of the functional Laplacian$^{8}$, so 
that the resulting equation will be simply a Volterra type-II linear 
integral equation, which can be investigated perturbatively.

All the points, mentioned above, will be the topic of the next Section.

The main results of the letter are summarized in the Conclusion.

\vspace{6mm}
{\large \bf 2. Loop equation for the partition function of the UCST and 
its investigation}

\vspace{3mm}

The partition function of the UCST, which is nothing else than the Wilson 
average in the Euclidean 4D compact QED, has the form$^{1,2}$

$$W[{\mathtt x}]=\left<\Phi\left(B_{\alpha\beta}\right)\right>$$
(we use the notations, emphasizing that the Wilson loop is a 
functional, defined on the loop space; here ${\mathtt x}\equiv x_\mu
(s), 0\le s\le 1,$ is an element of this space), 
where $\Phi\left(B_{\alpha\beta}\right)=\exp\left(i\int d\sigma_{\alpha
\beta}B_{\alpha\beta}\right),~$ $<...>\equiv\int DB_{\mu\rho} e^{-S}(...)$, 
and the action of the Kalb-Ramond field $B_{\mu\nu}$ reads as follows

$$S=\int d^4x\left[\frac{2}{3}\Lambda_0H_\mu arcsh\left(\frac{2}{3z\Lambda
_0^3}H_\mu\right)-z\Lambda_0^4\sqrt{1+\frac{4}{9z^2\Lambda_0^6}H_\mu^2}+
\frac{1}{4e^2}B_{\mu\nu}^2\right].\eqno (1)$$
Here $H_{\mu\nu\lambda}=\partial_\mu B_{\nu\lambda}+\partial_\nu B_{\lambda
\mu}+\partial_\lambda B_{\mu\nu}$ is a strength tensor of the  
field $B_{\mu\nu}, H_\mu=\varepsilon_{\mu\nu\alpha\beta} 
H_{\nu\alpha\beta}, \Lambda_0$ 
is a cutoff, which is necessary in 4D, $e$ is a dimensionless coupling 
constant, and $z\sim\exp\left(-\frac{const}{e^2}\right)$.

The equation of motion $\int DB_{\mu\rho}\frac{\delta}{\delta B_{\nu
\lambda}(x)}e^{-S}\Phi\left(B_{\alpha\beta}\right)=0$ has the form 

$$\left<\left[\frac{1}{2\Lambda^2}\left(\partial^2 B_{\nu\lambda}+
\partial_\mu\partial_\nu B_{\lambda\mu}+\partial_\mu\partial_\lambda 
B_{\mu\nu}\right)+\sqrt{1-\frac{z}{1536\Lambda^6}H_{\rho\sigma\zeta}^2}
\left(\frac{1}{2e^2}B_{\nu\lambda}-iT_{\nu\lambda}\right)\right]
\Phi\left(B_{\alpha\beta}\right)\right>=0,\eqno (2)$$
where $T_{\nu\lambda}(x)=
\int d^2\xi\varepsilon^{ab}\left(\partial_ax_\nu(\xi)\right)\left(
\partial_bx_\lambda(\xi)\right)\delta\left(x-x(\xi)\right)$ is the 
vorticity tensor current, and $\Lambda\equiv\frac{\Lambda_0}{4}
\sqrt{z}$. By virtue of Eq.(2) one gets the following 
loop equation

$$\Biggl[\frac{1}{2\Lambda^2}\left(\partial^{x(\sigma){\,}2}\frac{\delta}
{\delta\sigma_{\nu\lambda}(x(\sigma))}+\partial_\mu^x \partial_\nu^x
\frac{\delta}{\delta\sigma_{\lambda\mu}(x)}+\partial_\mu^x\partial_\lambda
^x\frac{\delta}{\delta\sigma_{\mu\nu}(x)}\right)+$$

$$+\sqrt{1+\frac{z}{512
\Lambda^6}\left[\left(\partial_\alpha^x\frac{\delta}{\delta\sigma
_{\beta\gamma}(x)}\right)^2+2\left(\partial_\alpha^x\frac{\delta}
{\delta\sigma_{\beta\gamma}(x)}\right)\left(\partial_\beta^x\frac
{\delta}{\delta\sigma_{\gamma\alpha}(x)}\right)\right]}\left(\frac{1}
{2e^2}\frac{\delta}{\delta\sigma_{\nu\lambda}(x)}+T_{\nu\lambda}
[{\mathtt x}]\right)\Biggr]W[{\mathtt x}]=0, \eqno (3)$$
where $\partial_\mu^{x(\sigma)}\equiv\int\limits_{\sigma-0}^{\sigma+0} 
d\sigma'\frac{\delta}{\delta x_\mu(\sigma')}$, and $T_{\nu\lambda}$ 
becomes a functional defined on the loop space.
 
Notice, that in the low-energy limit of the UCST action (1) 
takes the form 
$S=\int d^4 x\left(-\frac{1}{12\Lambda^2}H_{\mu\nu\lambda}^2+\frac{1}{4e^2}
B_{\mu\nu}^2\right)$, which yields the mass of the Kalb-Ramond field 
in this limit $m=\frac{\Lambda}{e}$, and therefore the validity of the 
low-energy 
approximation means that $\frac{H_{\mu\nu\lambda}^2}{z^2 e^6 m^6}\ll 1$. 
If this inequality holds true, the square root, standing on the L.H.S. 
of Eq.(3), replaces by unity. 

Let us now rewrite Eq.(3) in the momentum representation by making use 
of the functional Fourier transformation$^{6,7}$, which for an arbitrary 
functional $A[{\mathtt x}]$ is defined as $A[{\mathtt p}]=\int D{\mathtt x}
\exp\left(i\int\limits_0^1 d\sigma p_\mu\dot x_\mu\right)A[{\mathtt x}]$, 
where $D{\mathtt x}\equiv
{\cal D}{\mathtt x}\delta(x(0)-x(1))\delta(x(0)-const)$, and ${\cal D}
{\mathtt x}$ stands 
for the ordinary functional measure. Taking into account that$^{4}$ 
$\frac{\delta}{\delta\sigma_{\mu\nu}(x(\sigma))}=\int\limits_{-0}^{+0} 
d\tau\tau\frac{\delta^2}{\delta x_\mu\left(\sigma+\frac{1}{2}\tau\right)
\delta x_\nu\left(\sigma-\frac{1}{2}\tau\right)}$, we arrive at the 
following equation for the momentum Wilson loop

$$\left(\frac{e^2}{Q[{\mathtt p}]}\frac{\left(\Delta p\right)^2}{\Lambda^2}
{\bf P}_{\mu\rho, \lambda\nu}-{\bf 1}_{\mu\rho, \lambda\nu}\right)
\int\limits_{-0}^{+0} d\tau\tau\dot p_\nu\left(\sigma+\frac{1}{2}\tau
\right)\dot p_\lambda\left(\sigma-\frac{1}{2}\tau\right)W[{\mathtt p}]=
2e^2\int D{\mathtt p}'T_{\mu\rho}[{\mathtt p}-{\mathtt p}']
W[{\mathtt p}']. \eqno (4)$$
Here $\Delta p_\mu\equiv\Delta p_\mu(\sigma)=p_\mu(\sigma+0)-p_\mu
(\sigma-0)$ is the momentum discontinuity, 

$$Q\left[{\mathtt p}\right]
\equiv\Biggl(1+\frac{z}{512 \Lambda^6}\Delta p_\alpha\int\limits_{-0}^
{+0}d\tau\int\limits_{-0}^{+0}d\tau'\tau\tau'\dot p_\beta\left(\sigma+
\frac{1}{2}\tau\right)\dot p_\gamma\left(\sigma-\frac{1}{2}\tau\right)
\dot p_\gamma\left(\sigma-\frac{1}{2}\tau'\right)\Biggl(2\Delta p_\beta
\cdot$$

$$\cdot\dot p_\alpha\left(\sigma+\frac{1}{2}\tau'\right)
-\Delta p_\alpha
\dot p_\beta\left(\sigma+\frac{1}{2}\tau'\right)\Biggr)\Biggr)^
\frac{1}{2},~ {\bf P}_{\mu
\rho,\lambda\nu}=\frac{1}{2}\biggl({\cal P}_{\mu\lambda}{\cal P}_{\rho
\nu}-{\cal P}_{\mu\nu}{\cal P}_{\rho\lambda}\biggr),~ {\cal P}_{\mu\nu}
\equiv\delta_{\mu\nu}-\frac{\Delta p_\mu\Delta p_\nu}{\left(\Delta p
\right)^2},$$ 

$${\bf 1}_{\mu\rho, \lambda\nu}\equiv\frac{1}{2}\biggl(\delta
_{\mu\lambda}\delta_{\rho\nu}-\delta_{\mu\nu}\delta_{\rho\lambda}\biggr),$$ 
so that ${\bf P}_{\mu\rho, \lambda\nu}$ and $\left({\bf 1}-{\bf P}\right)_
{\mu\rho, \lambda\nu}$ are the projectors onto the transverse and 
longitudinal degrees of freedom of the momentum area derivative of the 
Wilson loop respectively, which satisfy the following relations 
${\bf P}^2={\bf P},~ \left({\bf 1}-{\bf P}\right)^2={\bf 1}-{\bf P}, 
~{\bf P}\left({\bf 1}-{\bf P}\right)=0$. The operator, standing in front 
of the integral  
on the L.H.S. of Eq.(4), can be easily inverted, which 
yields

$$\int\limits_{-0}^{+0}d\tau\tau\dot p_\nu\left(\sigma+\frac{1}{2}\tau
\right)\dot p_\lambda\left(\sigma-\frac{1}{2}\tau\right)W\left[
{\mathtt p}\right]=$$

$$=2e^2\left(\frac{\left(\Delta p\right)^2}{\left(
\Delta p\right)^2-\frac{Q\left[{\mathtt p}\right]}{e^2}\Lambda^2}
{\bf P}_{\lambda\nu,\mu\rho}-{\bf 1}_{\lambda\nu,\mu\rho}\right)
\int D{\mathtt p}'T_{\mu\rho}\left[{\mathtt p}-{\mathtt p}'\right]
W\left[{\mathtt p}'\right]. \eqno (5)$$
Eq.(5) splits into a pair of two independent equations for the transverse 
and longitudinal components of the momentum area derivative of the 
Wilson loop, which read correspondingly

$${\bf P}_{\mu\rho,\lambda\nu}\int\limits_{-0}^{+0}d\tau\tau\dot 
p_\nu\left(\sigma+\frac{1}{2}\tau\right)\dot p_\lambda\left(\sigma-
\frac{1}{2}\tau\right)W\left[{\mathtt p}\right]=$$

$$=2Q\left[{\mathtt p}
\right]\frac{\Lambda^2}{\left(\Delta p\right)^2-\frac{Q\left[{\mathtt p}
\right]}{e^2}\Lambda^2}{\bf P}_{\mu\rho, \lambda\nu}\int D{\mathtt p}'
T_{\lambda\nu}\left[{\mathtt p}-{\mathtt p}'\right]W\left[{\mathtt p}'
\right], \eqno (6)$$

$$\left({\bf 1}-{\bf P}\right)_{\mu\rho,\lambda\nu}\int\limits_{-0}^{+0}
d\tau\tau\dot p_\nu\left(\sigma+\frac{1}{2}\tau\right)\dot p_\lambda
\left(\sigma-\frac{1}{2}\tau\right)W\left[{\mathtt p}\right]=-2e^2
\left({\bf 1}-{\bf P}\right)_{\mu\rho,\lambda\nu}\int D{\mathtt p}'
T_{\lambda\nu}\left[{\mathtt p}-{\mathtt p}'\right]W\left[{\mathtt p}'
\right]. \eqno (7)$$
We see from Eq.(6) that when the momentum discontinuity satisfies the 
equation

$$\left(\Delta p\right)^2=\frac{Q\left[{\mathtt p}\right]}{e^2}\Lambda^2, 
\eqno (8)$$
the transverse component does not propagate. In the low-energy limit of 
the UCST the solution of Eq.(8) is simply $\left(\Delta p\right)^2=m^2$, 
which is obviously unphysical, since the momentum Wilson loop is known 
to have only a finite number of discontinuities$^{6}$, while such a 
solution would imply that a photon is emitted or absorbed at any point 
of the loop. Therefore in this limit the transverse component of the 
loop area derivative always propagates. 

Let us now further investigate Eq.(5). One can show that

$${\bf \Delta}W\left[{\mathtt x}\right]=\int D{\mathtt p} 
\exp\left(i\int\limits_0^1 d\sigma' x_\mu\dot p_\mu\right)
\int\limits_0^1 d\sigma\frac{\delta}
{\delta p_\nu(\sigma)}\Delta p_\lambda\int\limits_{-0}^{+0} d\tau\tau
\dot p_\nu\left(\sigma+\frac{1}{2}\tau\right)\dot p_\lambda\left(\sigma-
\frac{1}{2}\tau\right)W\left[{\mathtt p}\right], \eqno (9)$$
where ${\bf \Delta}=\int\limits_0^1 d\sigma\dot x_\nu\partial_\lambda^x
\frac{\delta}{\delta\sigma_{\lambda\nu}(x)}$ is the functional Laplacian, 
and $\frac{\delta}{\delta p_\nu(\sigma)}\equiv\frac{1}{2}\left(\frac{\delta}
{\delta p_\nu(\sigma+0)}+\frac{\delta}{\delta p_\nu(\sigma-0)}\right)$. 
Then, taking into account that $\dot p_\mu(\sigma)=\sum\limits_i^{}\Delta 
p_\mu(\sigma_i)\delta\left(\sigma-\sigma_i\right)$, where $\sigma_i$'s are 
the positions of the momentum discontinuities, and that when being applied 
to a functional without marked points ${\bf \Delta}$ may be rewritten as 
${\bf \Delta}=\int\limits_0^1 d\sigma\int\limits_{\sigma-0}^{\sigma+0}
d\sigma'\frac{\delta^2}{\delta x_\mu\left(\sigma\right)\delta x_\mu
\left(\sigma'\right)}$, we arrive by virtue of Eqs.(5) and (9) at the 
following equation for the momentum Wilson loop

$$\sum\limits_i^{}\left(\Delta p\left(\sigma_i\right)\right)^2W\left[
{\mathtt p}\right]=$$

$$=2e^2\int\limits_0^1 d\sigma\frac{\delta}{\delta 
p_\nu(\sigma)}\Delta p_\lambda\left({\bf 1}_{\lambda\nu,\mu\rho}-
\frac{\left(\Delta p\right)^2}{\left(\Delta p\right)^2-\frac{Q\left[
{\mathtt p}\right]}{e^2}\Lambda^2}{\bf P}_{\lambda\nu,\mu\rho}\right)
\int D{\mathtt p}'T_{\mu\rho}\left[{\mathtt p}-{\mathtt p}'\right]
W\left[{\mathtt p}'\right]. \eqno (10)$$
Eq.(10) describes the random motion of the momentum loop in such a way 
that the free propagation of the loop is described by the factor $\sum
\limits_i^{}\left(\Delta p\left(\sigma_i\right)\right)^2$ (see Ref.6), 
while the R.H.S. of Eq.(10) describes the interaction of the momentum loop 
with the functional vorticity tensor current.

To conclude with, we shall investigate loop equation,   
which one can derive by substituting Eq.(5) into Eq.(9), 
in the coordinate representation.  
To this end we shall make use of the method of inversion of the functional 
Laplacian, which was developed in$^{8}$ and applied in$^{9}$ to the 
solution of the Cauchy problem for the loop equation in turbulence. 
This procedure yields 
the following equation for the Wilson loop in the coordinate 
representation   

$$W\left[{\mathtt x}\right]=1+2e^2\int D{\mathtt p}\frac{1}
{\int\limits_0^1 d\sigma'\int\limits_0^1 d\sigma''\dot p_\alpha\left(
\sigma'\right)G\left(\sigma'-\sigma''\right)\dot p_\alpha\left(\sigma''
\right)}\cdot$$

$$\cdot\int\limits_0^1 d\sigma\Delta p_\lambda\left[2{\,}\frac{\int\limits
_0^1 d\sigma'\dot G\left(\sigma-\sigma'\right)\dot p_\nu\left(\sigma'\right)
\left(\exp\left(i\int\limits_0^1 d\sigma''x_\beta\dot p_\beta\right)-1
\right)}{\int\limits_0^1 d\sigma'\int\limits_0^1 d\sigma''\dot p_\gamma
\left(\sigma'\right)G\left(\sigma'-\sigma''\right)\dot p_\gamma\left(
\sigma''\right)}-i\dot x_\nu(\sigma)\exp\left(i\int\limits_0^1 d\sigma'
x_\beta\dot p_\beta\right)\right]\cdot$$

$$\cdot\left[\frac{\left(\Delta p\right)^2}
{\left(\Delta p\right)^2-\frac{Q\left[{\mathtt p}\right]}{e^2}\Lambda^2}
{\bf P}_{\lambda\nu,\mu\rho}-{\bf 1}_{\lambda\nu,\mu\rho}\right]
\int D{\mathtt y}\exp\left(-i\int\limits_0^1 d\sigma' y_\zeta\dot p_\zeta
\right)T_{\mu\rho}\left[{\mathtt y}\right]W\left[{\mathtt y}\right], 
\eqno (11)$$
where $G\left(\sigma-\sigma'\right)$ is a certain smearing function. 
Therefore we have reduced Eq.(3) to a simple Volterra type-II linear 
integral equation, which can be treated perturbatively (notice, that in 
the low-energy limit the coupling constant in Eq.(11) stands only in 
front of the integral operator, since $\frac{\Lambda^2}{e^2}$ replaces 
by $m^2$, and the iterative procedure of the solution of Eq.(11) 
linearizes).

\vspace{6mm}

{\large\bf 3. Conclusion}

\vspace{3mm}

In this letter we have derived and investigated loop equation for the 
partition function of the 4D Euclidean UCST. Our main result is the 
reduction of this equation, given by formula (3), to a simple Volterra 
type-II integral equation, which is given by formula (11).

Besides that, we have investigated the obtained loop equation in the 
momentum representation. Namely, we have inverted the functional 
differential operator, standing on the L.H.S. of Eq.(3), and obtained 
the equation for the momentum area derivative of the Wilson loop, which 
is given by formula (5). This equation decouples into a pair of two 
independent equations (6) and (7) for the transverse and longitudinal 
parts of this area derivative, and one can see that at the value of the 
momentum discontinuity, satisfying Eq.(8), the transverse part does not 
propagate. In the low-energy limit of the UCST Eq.(8) 
has only a spurious solution, and thus in this limit 
the transverse component always 
propagates. The obtained equation for the momentum 
area derivative of the Wilson loop has been also applied to the derivation 
of Eq.(10), which describes the random motion of the momentum loop in such 
a way, that its L.H.S. describes the free propagation of the loop, 
while its R.H.S. describes the interaction of the loop with the 
functional vorticity tensor current. 

\vspace{6mm}

{\large\bf 4. Acknowledgments}

\vspace{3mm}

I am deeply grateful to Professor Yu.M.Makeenko for bringing my attention 
to Refs.1 and 8 and a lot of useful discussions. I would also like to 
thank the theory group of the Quantum Field Theory Department of the 
Institut f\"ur Physik of the Humboldt-Universit\"at of Berlin for kind 
hospitality.

\newpage
{\large\bf References}

\vspace{3mm}
\noindent
1.~ A.M.Polyakov, preprint PUPT-1632 ({\it hep-th}/9607049).\\
2.~ M.C.Diamantini, F.Quevedo and C.A.Trugenberger, preprints 
CERN-TH/96-319, UGVA-DPT 1996/10-995 ({\it hep-th}/9612103).\\
3.~ P.Orland, {\it Nucl.Phys.} {\bf B428}, 221 (1994).\\
4.~ A.M.Polyakov, {\it Nucl.Phys.} {\bf B164}, 171 (1980).\\
5.~ Yu.M.Makeenko and A.A.Migdal, {\it Nucl.Phys.} {\bf B188}, 269 (1981), 
for a review see A.A.Migdal, {\it Phys.Rep.} {\bf 102}, 199 (1983).\\
6.~ A.A.Migdal, {\it Nucl.Phys.} {\bf B265 [FS15]}, 594 (1986).\\
7.~ A.A.Migdal, preprint PUPT-1509 ({\it hep-th}/9411100).\\
8.~ Yu.M.Makeenko, {\it Phys.Lett.} {\bf B212}, 221 (1988), preprints 
ITEP 88-50, ITEP 89-18; unpublished.\\
9.~ D.V.Antonov, {\it hep-th}/9612005 ({\it Mod.Phys.Lett.} {\bf A}, 
in press).\\
\end{document}